# Superconductivity at Pd/Bi$_2$Se$_3$ Interfaces Due to Self-Formed PdBiSe Interlayers

**Kaixuan Fan** [1,2,3,†], **Ze Hua** [4,†], **Siyao Gu** [1,2,3], **Peng Zhu** [5], **Guangtong Liu** [3], **Hechen Ren** [1,2], **Ruiwen Shao** [4,\*], **Zhiwei Wang** [5,\*], **Li Lu** [3] and **Fan Yang** [1,2,\*]

[1] Center for Joint Quantum Studies and Department of Physics, School of Science, Tianjin University, Tianjin 300354, China; fankaixuan@tju.edu.cn (K.F.); gusiyao@tju.edu.cn (S.G.); ren@tju.edu.cn (H.R.); fanyangphys@tju.edu.cn (F.Y.)
[2] Tianjin Key Laboratory of Low Dimensional Materials Physics and Preparing Technology, Department of Physics, Tianjin University, Tianjin 300354, China
[3] Beijing National Laboratory for Condensed Matter Physics, Institute of Physics, Chinese Academy of Sciences, Beijing 100190, China; gtliu@iphy.ac.cn (G.L.); lilu@iphy.ac.cn (L.L.)
[4] Beijing Advanced Innovation Center for Intelligent Robots and Systems, School of Medical Technology, Beijing Institute of Technology, Beijing, 100081, China; huae@bit.edu.cn (Z.H.); rwshao@bit.edu.cn (R.S.)
[5] Centre for Quantum Physics, Key Laboratory of Advanced Optoelectronic Quantum Architecture and Measurement (MOE), School of Physics, Beijing Institute of Technology, Beijing 100081, China; 3120215761@bit.edu.cn (P.Z.); zhiweiwang@bit.edu.cn (Z.W.)

† These authors contributed equally to this work.
\* Correspondence: rwshao@bit.edu.cn (R.S.); zhiweiwang@bit.edu.cn (Z.W.); fanyangphys@tju.edu.cn (F.Y.)

**Abstract:** Understanding the physical and chemical processes at the interface of metals and topological insulators is crucial for developing the next generation of topological quantum devices. Here we report the discovery of robust superconductivity in Pd/Bi$_2$Se$_3$ bilayers fabricated by sputtering Pd on the surface of Bi$_2$Se$_3$. Through transmission electron microscopy measurements, we identify that the observed interfacial superconductivity originates from the diffusion of Pd into Bi$_2$Se$_3$. In the diffusion region, Pd chemically reacts with Bi$_2$Se$_3$ and forms a layer of PdBiSe, a known superconductor with a bulk transition temperature of 1.5 K. Our work provides a method for introducing superconductivity into Bi$_2$Se$_3$, laying the foundation for developing sophisticated Bi$_2$Se$_3$-based topological devices.

## 1. Introduction

The non-trivial band topology of three-dimensional topological insulators (3D TIs) [1-3] makes it an advantageous platform for developing various types of topological devices [4]. For instance, a spinless topological superconductor can be artificially realized by introducing superconductivity via proximity effect into electronic states where spin degeneracy is lifted, such as the topological surface states of 3D TIs [5-6]. Such spinless superconductors are predicted to host Majorana fermions at their boundaries. When confined to zero-dimension, Majorana fermions develop into Majorana zero modes that obey non-Abelian statistics. Such quasiparticles are theoretically proposed as building blocks for fault-tolerant quantum computing [7-10]. Furthermore, the unique helical spin texture of topological surface states also renders 3D TIs promising for spintronic applications [10-13].

A deep understanding of the physical and chemical processes at metal/TI interfaces is essential for the design and fabrication of TI-based devices. Different functional interfaces can be achieved via appropriate selection of metals and precise control of the deposition conditions. Previously, it was reported that the diffusion of sputtered Pd into the 3D TI (Bi$_{1-x}$Sb$_x$)$_2$Te$_3$ leads to the self-formation of a superconducting PdTe$_2$ layer at the interface [14]. Similarly, signatures of superconductivity were also reported in annealed Pd/Bi$_2$Se$_3$ bilayer structures, but the origin of the superconductivity was unclear [15]. In this article, we report the observation of robust superconductivity in Pd/Bi$_2$Se$_3$ bilayers fabricated by sputtering Pd onto Bi$_2$Se$_3$. Although both Pd and Bi$_2$Se$_3$ are non-superconducting materials, the Pd/Bi$_2$Se$_3$ bilayer was found to exhibit a sharp superconducting transition at $T_c \approx 1.2$ K. Through atomically resolved structural analysis with transmission electron microscopy (TEM), it was determined that the observed superconductivity comes from a superconducting PdBiSe layer forming at the Pd/Bi$_2$Se$_3$ interface. Our work offers a new approach for introducing superconductivity into Bi$_2$Se$_3$ and paves the way for developing Bi$_2$Se$_3$-based hybrid superconducting devices.

## 2. Materials and Methods

### 2.1. Crystals Growth

High-quality single crystals of Bi$_2$Se$_3$ were grown using the melt method. Stoichiometric mixtures of Bi (99.9999% purity) and Se (99.999% purity) elements were melted in an evacuated quartz tube and then slowly cooled down to 550 °C over 50 hours. After that, the crystals were kept in the quartz tube at 550°C for 3 days and then cooled down to room temperature. Finally, large Bi$_2$Se$_3$ single crystals with shiny surfaces were obtained.

### 2.2. Device Fabrication

Bi$_2$Se$_3$ flakes were mechanically exfoliated from bulk single crystals and then transferred onto SiO$_2$ (300 nm)/Si substrates using polyethylene tapes, which produce less residual glue compared to commonly used Scotch tapes. Flakes with regular shapes and suitable thickness were selected for device fabrication. The resist pattern of Pd electrodes were prepared in a single step of electron-beam lithography (EBL). To remove possible residual resist and native oxide, the contacts areas were gently etched in Ar plasma for 40 seconds before the deposition of Pd. The etching power and Ar pressure were 2.6 W and 0.1 Pa, respectively. After etching, about 100 nm of Pd was deposited by magnetron sputtering with a power of 100 W and an argon pressure of 0.7 Pa. The lift-off of Pd was performed in acetone at 60°C.

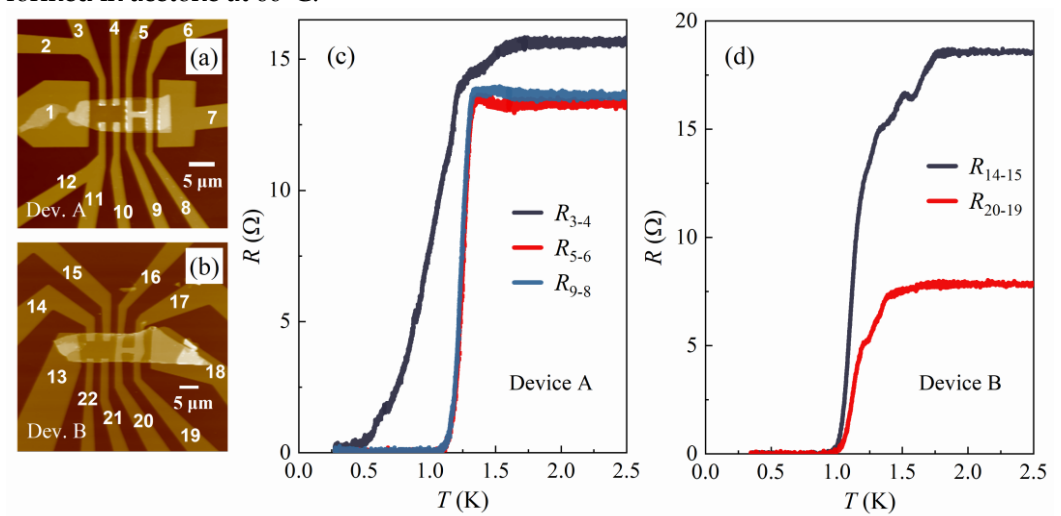

**Figure 1.** (**a**,**b**) AFM images of devices A and B, with electrode numbers indicated. The thickness of the Bi$_2$Se$_3$ flakes in devices A and B was measured to be 73 nm and 66 nm, respectively. (**c**,**d**) The $R(T)$ curves of devices A and B, measured with an excitation current of 50 nA.

After the fabrication, the devices were characterized using an atomic force microscope (AFM). In this paper, we present the data from two independent devices, labeled as devices A and B, respectively. The AFM images of these devices are shown in Figures 1(a) and 1(b), with all electrodes labeled accordingly.

*2.3. Transport Measurements*

Electron transport measurements were performed in a $^3$He cryostat with a base temperature of 270 mK in magnetic fields up to 14 T. The resistance was measured in a four-terminal geometry using standard lock-in technique.

*2.4 Structural Analysis*

For structural analysis using TEM, Pd/Bi$_2$Se$_3$ bilayer samples were prepared following the procedures described in Section 2.2.

The cross-sectional TEM lamellae of Pd/Bi$_2$Se$_3$ bilayer samples were prepared using a Thermo Fisher Helios G4 focused ion beam (FIB) system. For protection, about 1 μm of Pt was deposited on top of the bilayer sample before ion milling. After that, a cross-sectional lamella was cut from the Pt-capped sample using FIB and subsequently extracted and transferred onto the FIB-dedicated copper grid using a nanomanipulator for further thinning and polishing. Thinning of the lamella was performed using Ga+ ion beams with an acceleration voltage of 30 kV and a beam current of 230 pA. After thinning, the lamella was polished using Ga+ ions for a few minutes to minimize the surface damage induced by the thinning process. The acceleration voltage and beam current for the polishing process were 2 kV and 23 pA, respectively.

A Thermo Fisher Scientific Titan Themis Z 60-300 kV Electron Microscope with condenser lens and objective lens aberration correctors was used to image the atomic arrangement structure of Bi$_2$Se$_3$ and PdBiSe. The Energy-dispersive X-ray spectroscopy (EDX) mappings were gained through Bruker Super-X EDX detector. All high-angle annular dark-field imaging (HAADF) images were acquired at an atomic resolution of 80 pm with a beam current of 40 pA, a convergence semiangle of 21.5 mrad, and a collection semiangle snap of 80−379 mrad at 300 kV.

**3. Results and Discussion**

*3.1. Superconductivity of Pd/Bi$_2$Se$_3$ Bilayers*

The design of the devices [Figures 1(a) and (b)] allows two different configurations for four-probe resistance measurements.

The first configuration uses two independent Pd electrodes as voltage probes, such as electrodes #3 and #4 in device A [Figure 1(a)]. In this configuration, the measured resistance mostly comes from the Bi$_2$Se$_3$ flake between the Pd electrodes, and thus should typically take a finite value. Zero resistance can only be reached when the Bi$_2$Se$_3$ between the Pd electrodes becomes superconducting.

The second configuration, however, uses two voltage probes that are connected by a Pd "bridge" on top of the Bi$_2$Se$_3$ flake, such as electrodes #5 and #6 in device A [Figure 1(a)]. The advantage of this configuration is that it allows detecting the zero-resistance state caused by the superconductivity at the Pd/Bi$_2$Se$_3$ interface. In the absence of superconductivity, the measured resistance mainly comes from the Pd bridge and thus takes a finite value. However, once superconductivity occurs at the Pd/Bi$_2$Se$_3$ interface, the two voltage probes will be effectively shorted by the supercurrent. As a result, the measured four-probe resistance will drop to zero.

Surprisingly, at low temperatures, zero-resistance states were detected using both measurement configurations, as shown in Figures 1(c) and (d). According to the discussion above, these observations indicate that a superconducting phase forms not only directly beneath the Pd electrodes but also in the region within a few hundred nanometers from their edges, leading to a superconducting path connecting the two adjacent Pd electrodes. In device A, the $R(t)$ curve measured using independent Pd electrodes exhibits a lower $T_c$ compared to that measured across the Pd bridge, as shown in Figure 1(c), suggesting that the superconductivity developing between the Pd electrodes is weaker than that beneath the Pd layers. In contrast, in device B, the strength of superconductivity was found to be comparable across different regions in terms of $T_c$, which may be attributed to a longer lateral diffusion length of Pd in device B.

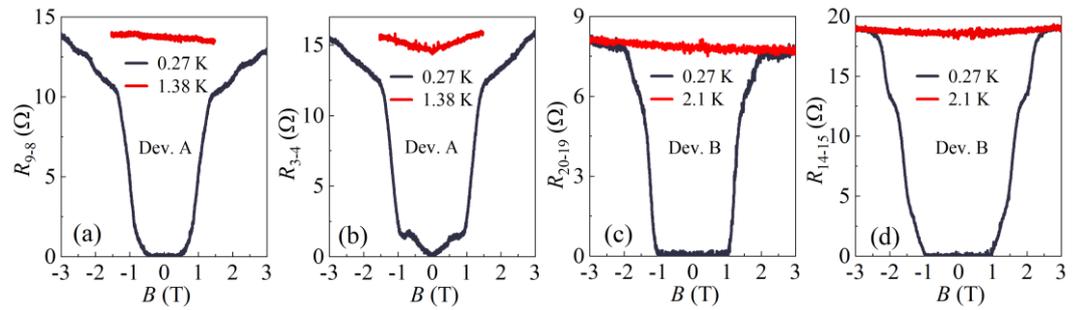

**Figure 2.** (**a**-**b**) Magnetic field dependence of the four-probe resistance (**a**) $R_{9\text{-}8}$ and (**b**) $R_{3\text{-}4}$ of device A, measured at 0.27 K and 1.38 K. (**c**,**d**) Magnetic field dependence of the four-probe resistance (**c**) $R_{20\text{-}19}$ and (**d**) $R_{14\text{-}15}$ of device B, measured at 0.33 K and 2.1 K. All curves were measured with an excitation current of 50 nA.

To obtain the upper critical magnetic field $B_{c2}$ of the observed superconducting phase, we measured the $R(B)$ curves of the devices at low temperatures, as plotted in Figure 2. Using 50% of the normal-state resistance as the criterion, the $B_{c2}$ values obtained in different regions of devices range between 1 T and 1.5 T, all below the Pauli-limiting field estimated using $B_P[\text{T}] \approx 1.84\, T_c[\text{K}]$ [16]. This indicates that the main mechanism for the suppression of superconductivity in magnetic fields is the orbital effect.

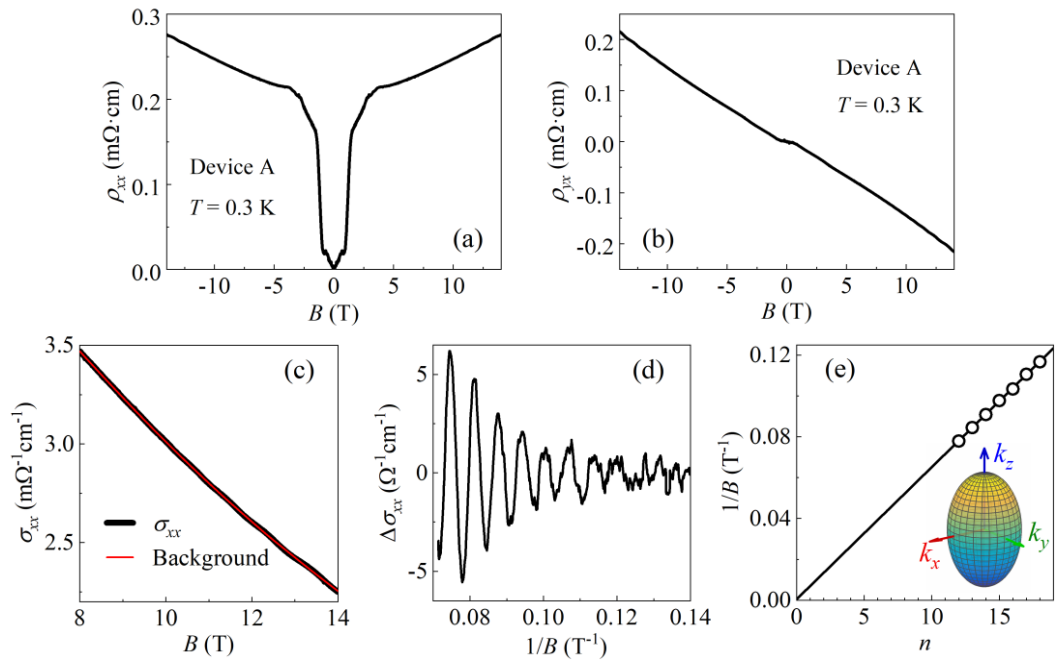

**Figure 3.** (**a**,**b**) The $\rho_{xx}(B)$ and $\rho_{yx}(B)$ curves of device A, where $\rho_{xx}(B)$ was measured using electrodes #3 and #4, and $\rho_{yx}(B)$ was measured using electrodes #3 and #12. High-field $\sigma_{xx}(B)$ data obtained using $\sigma_{xx} = \rho_{xx}/(\rho_{xx}^2 + \rho_{yx}^2)$, showing clear SdH oscillations. (**c**) The background of classical magnetoconductivity was obtained by smoothing the $\sigma_{xx}(B)$ curve, as shown by the red line in the figure. (**d**) The $\Delta\sigma_{xx}$ data obtained by subtracting the background, plotted against $1/B$.

(**e**) Landau fan diagram plotted using the minima of oscillations in (**d**). A linear fit to the data gives a zero intercept and a slope of 155 T. Inset: schematic diagram of the elliptical bulk Fermi surface of Bi$_2$Se$_3$.

In addition, it is noteworthy that the $B_{c2}$ values obtained in the region between independent Pd electrodes are comparable to those obtained beneath the Pd bridge, suggesting that the supercurrents flowing between adjacent Pd electrodes are not due to the Josephson effect, but rather originates from bulk superconducting states.

The coherence length of the superconducting states beneath the Pd bridge in device A was estimated to be $\xi(0) = \sqrt{\frac{\Phi_0}{2\pi B_{c2}(0)}} \approx 17$ nm, where $\Phi_0$ is the flux quantum. The value of $B_{c2}(0)$ was obtained by fitting the Ginzburg-Landau equation to the $B_{c2}(T)$ data, as shown in Figure S1 in the Supplementary Material. Such a $\xi$ is significantly smaller than both thickness of the Bi$_2$Se$_3$ flake and the distance between adjacent Pd electrodes.

*3.2. Shubnikov–de Haas Oscillations of Bi$_2$Se$_3$ Flakes*

Since zero-resistance states were detected between adjacent Pd electrodes, it is important to clarify whether the Bi$_2$Se$_3$ flake in this region remains intact. To investigate this issue, we measured the magnetoresistivity $\rho_{xx}(B)$ and Hall resistivity $\rho_{yx}(B)$ of the Bi$_2$Se$_3$ flake in device A using electrodes #3, #4, and #12. The results are plotted in Figures 3(a) and (b).

In high magnetic fields where superconductivity is suppressed, Shubnikov-de Haas (SdH) oscillations were observed in both $\rho_{xx}(B)$ and $\rho_{yx}(B)$ curves. To obtain more information, the $\sigma_{xx}(B)$ curve was calculated using $\sigma_{xx} = \rho_{xx}/(\rho_{xx}^2 + \rho_{yx}^2)$ in the high-field region, as shown in Figure 3(c). After subtracting the background of classical magnetoconductivity, clear signals of SdH oscillations were extracted, as plotted as a function of $1/B$ in Figure 3(d).

Figure 3(e) shows the Landau fan diagram of the SdH oscillations presented in Figure 3(d). The horizontal axis of the figure represents the Landau level index $n$, while the vertical axis represents the inverse of the magnetic field at which the oscillation minima occur. A linear fit to the data yields an oscillation frequency of $F = 155$ T and an intercept close to zero. The obtained frequency is in good agreement with previously reported values in Bi$_2$Se$_3$ in the literature [17-19]. Additionally, the zero intercept is also consistent with the zero Berry phase expected for bulk electrons in Bi$_2$Se$_3$, suggesting that the observed SdH oscillations originate from the bulk carriers of the Bi$_2$Se$_3$ flake.

The Fermi surface of Bi$_2$Se$_3$ is an ellipsoid with $k_c/k_{ab} = 1.62$ [20], as illustrated in the inset of Figure 3(e). The carrier density of Bi$_2$Se$_3$ is thus given by

$$n = \frac{1}{3\pi^2}\left(\frac{k_c}{k_{a,b}}\right)\left(\frac{2eF}{\hbar}\right)^{\frac{3}{2}}. \tag{1}$$

Substituting $F = 155$ T into Equation (1) leads to a carrier density of $n = 1.7 \times 10^{19}$ cm$^{-3}$, which is a typical value for bulk carriers of Bi$_2$Se$_3$ [17]. Together with the resistance value measured at $T = 2$ K, we obtain a bulk mobility of the Bi$_2$Se$_3$ flake $\mu = 1794$ cm$^2$V$^{-1}$s$^{-1}$.

The results above indicate that, from the perspective of electron transport, the crystal quality of the Bi$_2$Se$_3$ flake between the adjacent Pd electrodes is not affected. In the following section, we will further explore the reasons for these results through structural analysis.

*3.3. Formation of Superconducting PdBiSe*

To investigate the origin of the observed superconductivity, atomic-resolution elemental and structural analysis was performed using TEM. Technical details of TEM measurements are presented in Section 2.4.

Figure 4(a) shows a cross-sectional HAADF image taken near the edge of the Pd layer. The EDX elemental maps of this region [Figures 4(b)–(d)] provide clear evidence of Pd diffusion from the top Pd layer into the $Bi_2Se_3$ layer underneath. As shown in Figure 4(b), Pd fully penetrates through the $Bi_2Se_3$ flake in the regions directly beneath the Pd layer and within about 300 nm from the edge. Meanwhile, in areas about 300 nm to 650 nm away from the edge of the Pd layer, Pd continues to diffuse outward along the top and bottom surfaces of the $Bi_2Se_3$ flake, but no distribution of Pd was observed in the center of the $Bi_2Se_3$ flake.

To investigate whether the observed superconductivity in $Bi_2Se_3$/Pd bilayers is caused by Pd diffusion, we imaged both the Pd-free and Pd-rich regions in the $Bi_2Se_3$ flake using TEM. In Pd-free regions far from the Pd/$Bi_2Se_3$ interface, the crystal structure of $Bi_2Se_3$ remains intact, exhibiting clear stacking of Se-Bi-Se-Bi-Se quintuple layers, as shown in Figure 5(a). However, in Pd-rich regions, the sample exhibits a crystal structure distinctly different from that of $Bi_2Se_3$, as shown in the region enclosed by the red dashed box in Figure 5(b).

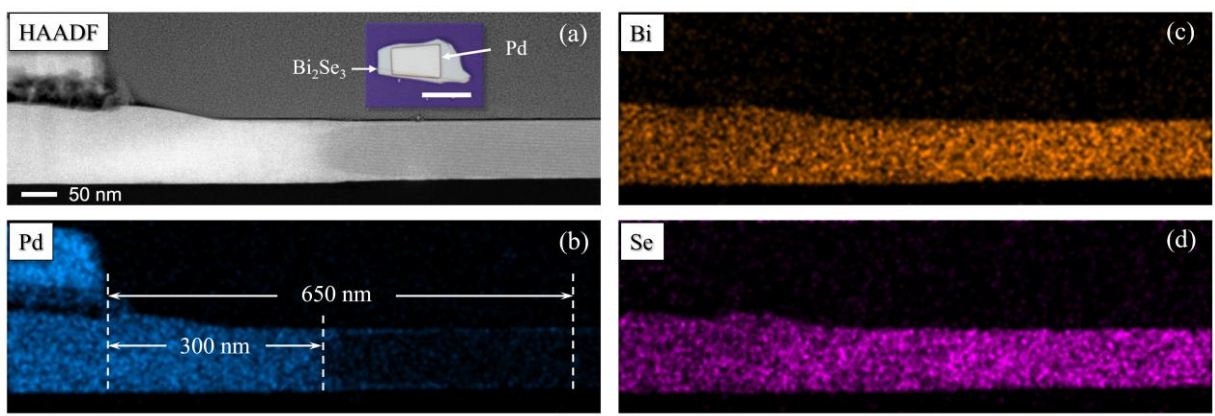

**Figure 4**. (**a**) Cross-sectional HAADF image of a Pd/$Bi_2Se_3$ bilayer, taken at the edge of the Pd layer. Inset: optical photo of an as-fabricated Pd/$Bi_2Se_3$ bilayer sample, with a scale bar of 10 μm. (**b**–**d**) EDX Elemental maps of (**b**) Pd, (**c**) Bi and (**d**) Se of the Pd/$Bi_2Se_3$ bilayer.

The difference in lattice structures of these two regions is also evident in their 2D fast Fourier transform (FFT) patterns, as shown in Figures 5(c) and (d). The square-shaped FFT pattern of the Pd-rich region [Figure 5(c)] indicates that the crystalline compound found in this area has a cubic lattice structure, with a lattice constant of 0.63 nm. After a comprehensive comparison of crystal structure and lattice constant with known materials in the database, the observed compound was identified as PdBiSe.

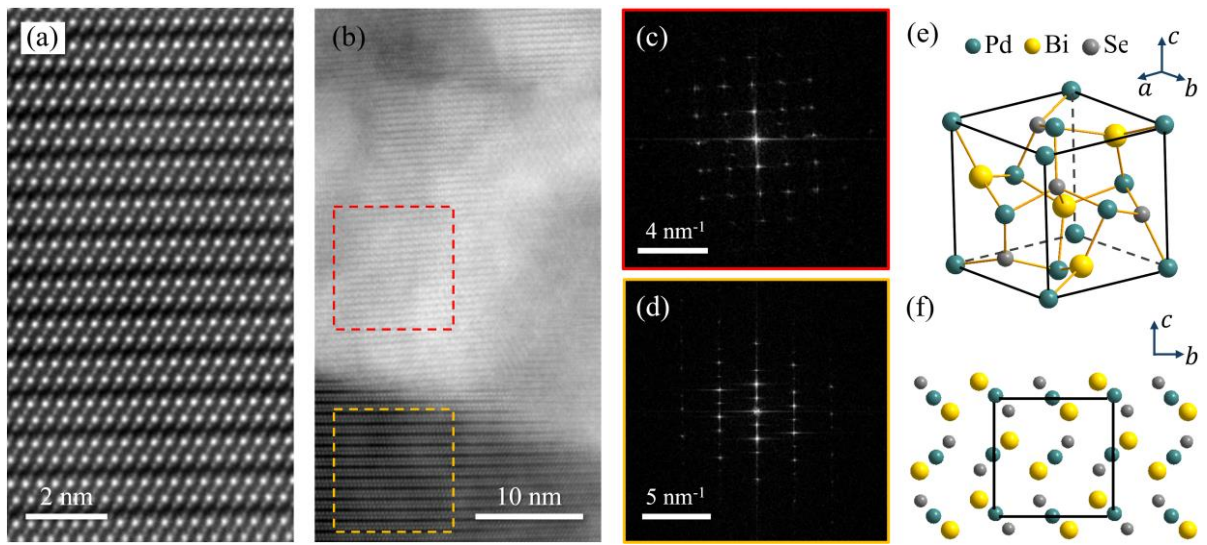

**Figure 5**. (**a**) Atomic-resolution HAADF image of the $Bi_2Se_3$ layer far from the $Pd/Bi_2Se_3$ interface, showing clear stacking of Se-Bi-Se-Bi-Se quintuple layers. (**b**) HAADF image at the $Pd/Bi_2Se_3$ interface. A crystalline phase was found to form in the Pd-rich area. (**c**,**d**) 2D FFT patterns of the areas indicated by the dashed (**c**) red and (**d**) yellow boxes in (**b**). The lattice constant extracted from the data in (**c**) is 0.63 nm, fairly close to that of cubic PdBiSe. (**e**) The unit cell of PdBiSe, with a lattice constant of 0.64 nm. (**f**) Schematics of the PdBiSe crystal structure. The black box represents the unit cell.

PdBiSe is a superconductor with a noncentrosymmetric cubic structure, as illustrated in Figures 5(e) and (f). The lattice constant of PdBiSe is 0.64 nm, which is very close to that of the crystalline compound found in the Pd-rich regions of the $Pd/Bi_2Se_3$ bilayers [21]. Although PdBiSe is not a van der Waals material, it exhibits a layered structure when viewed along the $a$-axis, as illustrated in Figure 5(f). Such a layered structure is consistent with the layer-like patterns observed in the Pd-rich region in Figure 5(b).

The bulk $T_c$ of PdBiSe is approximately 1.5 K, slightly higher than the maximum $T_c$ of 1.25 K observed in our devices [Figure 1(c)]. Since the $T_c$ of a superconductor is often influenced by various factors such as size and crystal quality, such a small difference in $T_c$ is not surprising.

## 4. Further Discussion

In our early work, we (F.Y., G.L., and L.L) used Pd electrodes as normal-metal contacts to probe the proximity-induced superconductivity in $Bi_2Se_3$ [22], unaware that a superconducting compound could form at the $Pd/Bi_2Se_3$ interface. With the knowledge gained from this study, a reinterpretation of the results in Ref. [22] is necessary.

## 5. Conclusions

We discovered that sputtering Pd onto $Bi_2Se_3$ flakes leads to the formation of a crystalline PdBiSe layer with a superconducting transition temperature of $T_c \approx 1.2$ K. These findings not only deepen the understanding of the physical and chemical processes at the $Pd/Bi_2Se_3$ interface but also provide a practical approach for introducing superconductivity into the topological insulator $Bi_2Se_3$, and therefore will be helpful for the development of $Bi_2Se_3$-based hybrid superconducting devices.

**Author Contributions:** F.Y. conceived and supervised the project; P.Z. and Z.W. grew the $Bi_2Se_3$ single crystals; K.F., S.G., G.L., and L.L. fabricated the devices and performed low-temperature transport measurements; Z.H and R.S. performed TEM measurements; K.F., Z.H., R.S., H.R. and F.Y. analyzed the data; F.Y. wrote the manuscript with the input of all authors.

**Funding:** This research was funded by National Natural Science Foundation of China, Grants No. 11904259 and No. Z210006; Beijing National Laboratory for Condensed Matter Physics, Grant No. 2023BNLCMPKF007. H.R. acknowledges funding support by the National Natural Science Foundation of China.

**Institutional Review Board Statement:** Not applicable.

**Informed Consent Statement:** Not applicable.

**Data Availability Statement:** The data presented in this study are available on request from the corresponding authors.

**Acknowledgments:** The authors thank the Analsis & Testing Center, Beijing Institute of Technology for the use of scanning TEM and FIB. The authors thank Xiaohui Song for the help on device fabrication.

**Conflicts of Interest:** The authors declare no conflict of interest.

# Supplementary Materials for "Superconductivity at Pd/Bi₂Se₃ Interfaces Due to Self-Formed PdBiSe Interlayers"

**Kaixuan Fan** [1,2,3,†], **Ze Hua** [4,†], **Siyao Gu** [1,2,3], **Peng Zhu** [5], **Guangtong Liu** [3], **Hechen Ren** [1,2], **Ruiwen Shao** [4,*], **Zhiwei Wang** [5,*], **Li Lu** [3] **and Fan Yang** [1,2,*]

[1] Center for Joint Quantum Studies and Department of Physics, School of Science, Tianjin University, Tianjin 300354, China; fankaixuan@tju.edu.cn (K.F.); gusiyao@tju.edu.cn (S.G.); ren@tju.edu.cn (H.R.); fanyangphys@tju.edu.cn (F.Y.)
[2] Tianjin Key Laboratory of Low Dimensional Materials Physics and Preparing Technology, Department of Physics, Tianjin University, Tianjin 300354, China
[3] Beijing National Laboratory for Condensed Matter Physics, Institute of Physics, Chinese Academy of Sciences, Beijing 100190, China; gtliu@iphy.ac.cn (G.L.); lilu@iphy.ac.cn (L.L.)
[4] Beijing Advanced Innovation Center for Intelligent Robots and Systems, School of Medical Technology, Beijing Institute of Technology, Beijing, 100081, China; huaze@bit.edu.cn (Z.H.); rwshao@bit.edu.cn (R.S.)
[5] Centre for Quantum Physics, Key Laboratory of Advanced Optoelectronic Quantum Architecture and Measurement (MOE), School of Physics, Beijing Institute of Technology, Beijing 100081, China; 3120215761@bit.edu.cn (P.Z.); zhiweiwang@bit.edu.cn (Z.W.)

† These authors contributed equally to this work.
\* Correspondence: rwshao@bit.edu.cn (R.S.); zhiweiwang@bit.edu.cn (Z.W.); fanyangphys@tju.edu.cn (F.Y.)

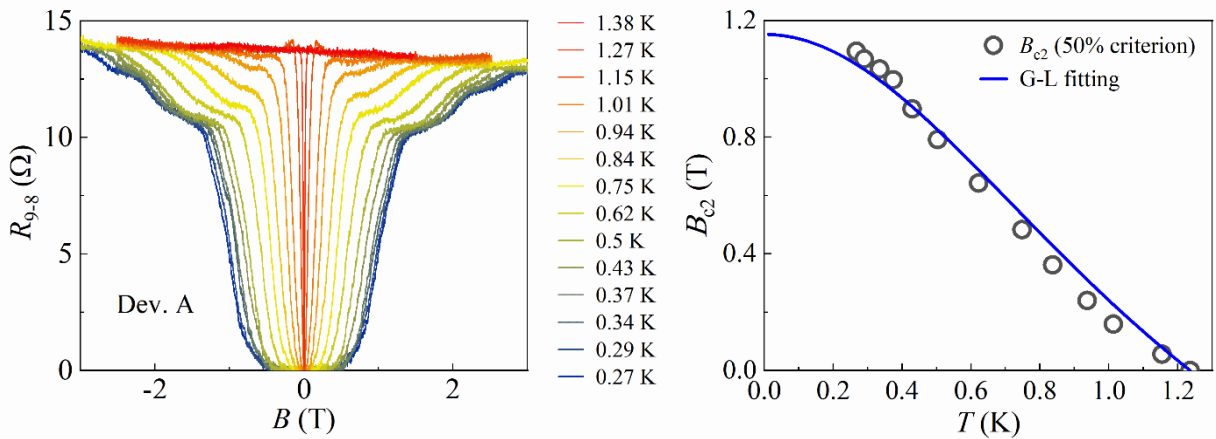

**Figure S1**. (a) The $R_{9\text{-}8}(B)$ curves of device A, measured at various temperatures. (b) The $B_{c2}(T)$ data extracted from the curves in (a). Here, the $B_{c2}$ is defined as the magnetic field at which $R_{9\text{-}8}$ reaches 50% of its normal-state value. The blue line represents the best fit to the $B_{c2}(T)$ data using the Ginzburg-Landau equation $B_{c2}(T) = B_{c2}(0)\left[\frac{1-(T/T_c)^2}{1+(T/T_c)^2}\right]$, which yields $B_{c2}(0) = 1.15$ T.